\documentclass[sn-mathphys,Numbered,iicol]{sn-jnl}

\usepackage{graphicx}%
\usepackage{multirow}%
\usepackage{amsmath,amssymb,amsfonts}%
\usepackage{amsthm}%
\usepackage{mathrsfs}%
\usepackage[title]{appendix}%
\usepackage{xcolor}%
\usepackage{textcomp}%
\usepackage{manyfoot}%
\usepackage{booktabs}%
\usepackage{algorithm}%
\usepackage{algorithmicx}%
\usepackage{algpseudocode}%
\usepackage{listings}%
\usepackage{subcaption}
\newcommand{\sech}{\mathrm{sech}}

\theoremstyle{thmstyleone}%

\theoremstyle{thmstyletwo}%

\theoremstyle{thmstylethree}%

\begin{document}

\title[Detecting Delamination via Nonlinear Wave Scattering in a Bonded Elastic Bar]{Detecting Delamination via Nonlinear Wave Scattering in a Bonded Elastic Bar}

\author[1]{\fnm{Jagdeep S.} \sur{Tamber}}

\author[1]{\fnm{David J.} \sur{Chappell}}

\author[1]{\fnm{Jack C.} \sur{Poore}}

\author*[1]{\fnm{Matt R.} \sur{Tranter}}\email{Matt.Tranter@ntu.ac.uk}

\affil[1]{\orgdiv{Department of Physics and Mathematics}, \orgname{Nottingham Trent University}, \orgaddress{\postcode{NG11 8NS}, \country{United Kingdom}}}

\abstract{In this paper we examine the effect of delamination on wave scattering, with the aim of creating a control measure for layered waveguides of various bonding types. Previous works have considered specific widths of solitary waves for the simulations, without analysing the effect of changing the soliton parameters. We consider two  multi-layered structures: one containing delamination `sandwiched' by perfect bonding and one containing delamination but `sandwiched' by soft bonding. These structures are modelled by coupled Boussinesq-type equations. Matched asymptotic multiple-scale expansions lead to coupled Ostrovsky equations in soft bonded regions and Korteweg--De Vries equations in the perfectly bonded and delaminated region. We use the Inverse Scattering Transform to predict the behaviour in the delaminated regions.

In both cases, numerical analysis shows that we can predict the delamination length by changes in the wave structure, and that these changes depend upon the Full Width at Half Magnitude (FWHM) of the incident soliton. In the case of perfect bonding, we derive a theoretical prediction for the change and confirm this numerically. For the soft bonding case, we numerically identify a similar relationship using the change in amplitude. Therefore we only need to compute one curve to determine the behaviour for any incident solitary wave, creating a framework for designing measurement campaigns for rigorously testing the integrity of layered structures.}




\maketitle

\section{Introduction}
\label{intro}
Solitary waves are of significant interest from both a mathematical perspective as well as in physical and engineering applications. They often arise as solutions to nonlinear equations such as the KdV equation (and its extensions) in shallow water \cite{Whitham74, Ablowitz81, Drazin89, Karczewska14}, the Benjamin-Ono equation for internal waves of stratified fluids \cite{Benjamin67, Ono75}, the nonlinear Schr\"{o}dinger equation in optics \cite{Stegeman99, Yang10}, and flexural waves in the beam equation \cite{Champneys00}, to name a few. From a purely mathematical perspective, there are many studies into the existence and behaviour of solitons, for example as solutions to the Boussinesq equation  \cite{Gao17, Hu21}. Boussinesq-type equations are of interest in this study, in the context of solid mechanics. It is well-known that they can describe long longitudinal bulk strain waves in elastic waveguides, such as rods and metal plates (see e.g. \cite{Samsonov01, Porubov03, Peake06, Peets17, Andrianov19, Garbuzov19}). Practical experiments have confirmed that longitudinal bulk strain solitons exist in these waveguides, which validated theoretical findings \cite{Dreiden08, Dreiden11, Dreiden12, Dreiden14}. 

Indeed, layered waveguides with bonding between the layers can be modelled by the so-called ``doubly dispersive equation'' (DDE), which can be derived using nonlinear elasticity theory for long longitudinal waves in a bar of rectangular cross-section \cite{Samsonov01, Porubov03}. The DDE for a bar of rectangular cross-section $\sigma = 2a \times 2b$ has the form
\begin{equation}
    f_{tt}-c^2f_{xx}=\frac{\beta}{2\rho}(f^2)_{xx} +\frac{Jv^2}{\sigma}(f_{tt}-c_1^2f_{xx})_{xx},
    \label{DDE}
\end{equation}
where 
\begin{align}
\beta&=3E+2l(1-2v)^3+4m(1+v)^2(1-2v)+6nv^2, \notag \\
c&=\sqrt{\frac{E}{\rho}}, \quad c_1=\frac{c}{\sqrt{2(1+v)}}, \quad J=\frac{4ab(a^2+b^2)}{3},
	\label{DDEParam}
\end{align}
$\rho$ is the density, $E$ is the Young's modulus, $v$ is the Poisson's ratio, while $l$, $m$, $n$ are the  Murnaghan's moduli, and $a$ and $b$ are geometric and physical parameters.

The case when the interlayer bonding is missing over part of the structure, known as delamination, is important for a wide range of applications in non-destructive testing for structural damage in the multi-layer beam-like structures found throughout civil and mechanical engineering.  The governing mathematical model then takes the form of a scattering problem and, for a perfectly bonded waveguide (represented in experiments by cyanoacrylate), we find a series of Boussinesq equations, with continuity conditions on the interface between sections \cite{Khusnutdinova08}. Incident solitons fission into multiple solitons with dispersive radiation, agreeing with theoretical predictions \cite{Khusnutdinova08}, numerical simulations \cite{Khusnutdinova15, Tranter19} and experimental observations \cite{Dreiden12,Dreiden10, Dreiden10Strain}.  

In the case of an imperfect ``soft'' bonding (represented in experiments by polychloroprene), a model based upon a series of anharmonic coupled dipoles can be used to derive coupled regularised Boussinesq (cRB) equations to model long nonlinear longitudinal bulk strain waves in a bi-layer, assuming sufficiently soft bonding \cite{Khusnutdinova09}. In this case, when the materials in the layers are sufficiently close, an incident solitary wave evolves in the bonded region into a solitary wave with a one-sided, co-propagating oscillatory tail, known as a radiating solitary wave. In the delaminated regions, the solitary wave detaches from its tail and this can be used as a measure of delamination \cite{Khusnutdinova17}.  More recently, in the limiting case of a semi-infinite delamination, we find the emergence of Ostrovsky wave packets in bonded regions \cite{Tamber22}. The Ostrovsky equation originally arose in the context of shallow-water waves, where the rotation of the Earth is considered \cite{Ostrovsky78}, and the evolution of wave packets generated from an initial pulse has been extensively studied \cite{Grimshaw08}.

In this paper we aim to use theoretical predictions and numerical simulations to establish a prediction for the delamination length based upon changes in the wave during its propagation.  We will consider a range of initial conditions by varying the Full Width at Half Magnitude (FWHM) of the incident wave, whereas previous studies have only considered a single fixed incident soliton \cite{Khusnutdinova15, Khusnutdinova17}.  Our aim is to find a relationship between the generated delamination curves for different values of FWHM, so that only one curve needs to be computed, significantly reducing the computation time.  This wider range of predictions allows for the design of measurement campaigns for detecting and measuring delamination in layered waveguides.  We will consider a multi-layered symmetric structure with perfect bonding, as well as a two-layered structure with soft bonding. In both cases, we will consider delamination `sandwiched' by bonding. These structures are illustrated in figures \ref{fig:PerfectBond} and \ref{fig:SBFig}, and are inspired by an existing experimental set-up \cite{Dreiden12}.

The paper is structured as follows. In Section \ref{sec:Setup} we introduce the equations describing longitudinal wave propagation in both the perfectly bonded and soft bonded cases. We also introduce the weakly-nonlinear solution for the perfectly bonded case so that we can create a measure of the delamination length using theoretical predictions. In Section \ref{sec:Numerical}, we begin by illustrating the evolution of incident solitary waves in the cases of both a perfectly bonded waveguide and a soft bonded waveguide.  Next, for the perfectly bonded case, we use theoretical predictions to determine the length of the delamination region for a variety of incident solitary waves, tested via numerical simulations and measuring the amplitude of the transmitted wave with reference to the incident wave. This gives rise to a relationship between FWHM of the incident soliton and the delamination length, allowing for efficient computation for other incident waves. A similar result is also found for the soft bonded case, analysing the decrease in amplitude after the wave propagates through a delaminated region. The theoretical prediction is difficult in this case, so we rely on numerical observations and instruction from the previous case. In Section \ref{sec:Conc} we conclude our discussions.

\section{Problem set-up}
\label{sec:Setup}
\subsection{Perfectly bonded case}
\begin{figure}[htbp] 
   \centering
 	\includegraphics[width=0.9\linewidth]{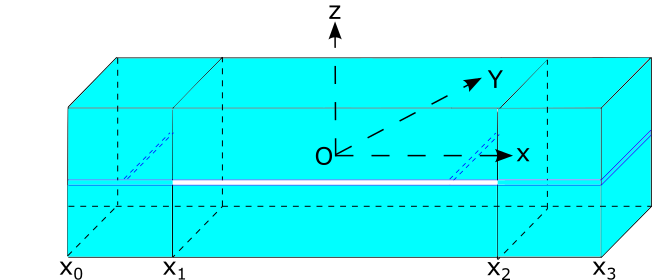} 
    \vspace{4ex}
  \caption{Bi-layer structure with an initial perfect bonded region for $x_0<x<x_1$, a delaminated region for $x_1<x<x_2$ and a perfect bonded region for  $x_2<x<x_3$. We assume that the materials in both layers are identical.} 
  \label{fig:PerfectBond} 
\end{figure}
We consider the scattering of a long longitudinal strain solitary wave in a perfectly bonded layered bar with delamination in the centre, as illustrated in Figure \ref{fig:PerfectBond}. Note that we only illustrate two layers in this figure, but the setup can accommodate any number of symmetric layers, as we assume that the materials of the layers are identical and that the bonding is the same between all layers. This problem is described by the regularized non-dimensional Boussinesq equations \cite{Khusnutdinova08}
\begin{align}
    u^{(i)}_{tt} - c_i^2 u^{(i)}_{xx} & = \varepsilon \left[-12 \alpha_i u^{(i)}_{x} u^{(i)}_{xx} + 2\beta_i u^{(i)}_{ttxx} \right],
    \label{PBEq}
\end{align}
where $i=1,3$ represent the perfect bonded regions and $i=2$ represents the delaminated region. We have the coefficients $\alpha_i$, $\beta_i$ and $c_i$, which can theoretically vary between sections (representing a waveguide with different materials in each section), but for our purposes we will assume that the sections are of one and the same material. The parameter $\varepsilon$ is the small wave parameter. The Boussinesq equations are complemented with continuity conditions, namely continuity of longitudinal displacement 
\begin{equation}
    \left. u^{(i)} \right|_{x=x_i} = \left. u^{(i+1)}\right|_{x=x_i},
    \label{PBContDisp}
\end{equation}
and continuity of normal stress
\begin{equation}
\sigma^{(i)}|_{x=x_i} = \sigma^{(i+1)}|_{x=x_i},
    \label{PBContStress}
\end{equation}
where $\sigma^{(i)}$ is defined by our original equation \eqref{PBEq} when written in the form
\begin{equation*}
	u_{tt}^{(i)} = \frac{\mathrm{d} \sigma^{(i)}}{\mathrm{d} x}.
\end{equation*}
We consider $\alpha_i=1$ for all $i$, $\beta_{1,3} = 1$ and 
\begin{equation}
    \beta_2(n,k) = \frac{n^2+k^2}{n^2(1+k^2)},
    \label{beta}
\end{equation}
where $n$ represents the number of layers in the structure and $k$ is defined by the geometry of the waveguide. Referring to Figure \ref{fig:PerfectBond}, the cross section has width $2a$ and the height of each layer is $2b/n$. In terms of these values, as there are two layers, $n = 2$ and $k = b/a$. In our numerical simulations we will consider various $n$ and $k$ values.

\subsubsection{Weakly-nonlinear solution}
In order to find theoretical predictions for the evolution of the solitary waves, we construct a weakly-nonlinear solution and use theoretical results for the derived equations. For brevity, we only provide a summary of the results below, more details can be found in Refs.~\cite{Khusnutdinova15, Tranter19, Khusnutdinova17, Tamber22}. We seek a weakly-nonlinear solution for the strains $f^{(i)} = u_x^{(i)}$ of the form
\begin{align}
     f^{(i)} &= T^{(i)}(\xi,X) + R^{(i)}(\eta,X) + \varepsilon P^{(i)}(\xi,\eta,X) \notag \\
     &~~~+ \mathcal{O}\left(\varepsilon^2\right),
     \label{PBWNL}
\end{align}
where $\xi = x - c_i t$, $\eta = x + c_i t$ and $X = \varepsilon x$. Substituting the respective weakly-nonlinear solutions into the differentiated form of \eqref{PBEq}, then applying space-averaging (see \cite{Khusnutdinova15, Khusnutdinova17, Tamber22}) yields the leading order solutions
\begin{align}
	T^{(i)}_{X} - 6 \frac{\alpha_{i}}{c_i^2} T^{(i)}T^{(i)}_{\xi} + \beta_{i} T^{(i)}_{\xi \xi \xi} &= 0, \label{TKdV} \\ 
    R^{(i)}_{X} - 6 \frac{\alpha_{i}}{c_i^2} R^{(i)} R^{(i)}_{\eta} + \beta_{i} R^{(i)}_{\eta \eta \eta} &= 0. \label{RKdV}
\end{align}
To determine ``initial conditions'' for the equations derived in each section, we substitute \eqref{PBWNL} into the continuity conditions \eqref{PBContDisp} -- \eqref{PBContStress} to find values for $T$ and $R$ at the interface, in terms of the previous transmitted wave. This gives rise to transmission and reflection coefficients in terms of $c_i$. As we have assumed that the waveguide is one and the same material (so $c_i = 1$ for all $i$) we have full transmission and no reflection.

\subsubsection{Theoretical predictions}
We can rewrite the transmitted wave equation \eqref{TKdV} in the form
\begin{equation}
    U_{\tau} - 6 U U_{\chi} + U_{\chi \chi \chi} = 0, \quad U |_{\tau = 0} = U_0 ( \chi ).
    \label{UKdV}
\end{equation}
For a sufficiently rapidly decaying initial condition $U_0( \chi )$ on the infinite line, the solution to \eqref{UKdV} is related to the spectral problem for the Schr\"{o}dinger equation
\begin{equation}
	\Psi_{\chi \chi} + [ \lambda - U_0 ( \chi ) ] \Psi = 0, 
	\label{SchroEq}
\end{equation}
where $\lambda$ is the spectral parameter.  Finding the evolution of the scattering data for the discrete and continuous spectra and using these to reconstruct the solution to the KdV equation is known as the Inverse Scattering Transform (IST) \cite{GGKM67}. We can use the results from the IST to create theoretical predictions for the solitons in the delaminated region, as well as in the second bonded region.

We assume that there is an incident soliton in the first region, which is a travelling wave solution and thus will move in time but retains its shape. To illustrate the theoretical predictions we consider the second region, where we have $\beta_2$ defined as in \eqref{beta} and the initial condition for \eqref{UKdV} in this region then takes the form 
\begin{equation}
    U_0(\chi) = -A\sech^2 \left(\frac{\chi}{l}\right), \quad A = \frac{v}{2\beta_{2}}, \quad l = \frac{2}{\sqrt{v}}.
    \label{KdVIC}
\end{equation}
In this case the solution will consist of either one soliton, or a series of solitons, characterised by eigenvalues in the discrete spectrum, and accompanying dispersive radiation determined by the continuous spectrum. In some cases we may see the fission of the initial soliton, which is when more than one soliton is generated, in particular when $\beta_{2} \neq 1$.

The discrete eigenvalues of \eqref{SchroEq} take the form $\lambda = -k_n^2$, where 
\begin{equation}
	k_n = \frac{1}{2l} \left[ \sqrt{1 + 4Al^2} - (2n - 1) \right],
\end{equation}
for $n=1,2,\dots,N$. Recalling \eqref{beta}, the number of solitons, $N$, generated in the delaminated region is given by the largest integer satisfying the inequality
\begin{equation}
    N < \frac{1}{2} \left( \sqrt{1 + \frac{8}{\beta_{2}}} + 1 \right).
    \label{NSol}
\end{equation} 
We can see from \eqref{NSol} that, for $\beta = 1$ we will have one soliton, while for $\beta < 1$, we will have more than one soliton and as $\beta$ becomes smaller, more solitons will be generated. This corresponds to either an increase in layers in the waveguide, or a change in geometry.  
As $\tau \rightarrow \infty$, the solution will evolve into a train of solitary waves, ordered by their heights, propagating to the right and some dispersive radiation (a dispersive wave train) propagating to the left (in the moving reference frame), i.e.
\begin{align}
    U(\chi,\tau) &\sim -\sum^N_{n=1}2k_n^2\textrm{sech}^2(k_n(\chi -4k_n^2\tau-\chi_n)) \notag \\
    &~~~+ \textrm{radiation},
\end{align}
where $\chi_n$ is the phase shift. In the context of our problem, if there is an infinite delamination then the solitons will separate and rank order, while for finite delamination the solitons will only separate for a large delamination. This allows us to create a measure of the delamination length, by comparing the measured signal at the end of the bar to the theoretical prediction.

We introduce the incident solitary wave for $T^{(1)}$, the exact travelling wave solution of \eqref{TKdV}, as 
\begin{equation}
    T^{(1)}(\xi, X) = -\frac{v}{2} \sech^2 \left(\frac{\sqrt{v}}{2} \left(\xi - v X \right) \right),
    \label{T1IC}
\end{equation}
where $v$ is the phase speed. Solitary waves are often measured in experiments in terms of their Full Width at Half Magnitude (FWHM), so we rewrite this as
\begin{equation}
    -\frac{v}{2} \sech^2 \left(\frac{\sqrt{v}}{4} \mathrm{FWHM} \right) = -\frac{v}{4},
\end{equation}
and hence we obtain
\begin{equation}
	v = \left( \frac{4}{\mathrm{FWHM}} \cosh^{-1}(\sqrt{2})\right)^2.
    \label{FWHM}
\end{equation}
This allows us to generalise the FWHM based measure to any size of incident solitary wave.

\subsection{Imperfect bonding case}
\label{sec:ImperfBond}
\begin{figure}[htbp] 
   \centering
    \includegraphics[width=0.9\linewidth]{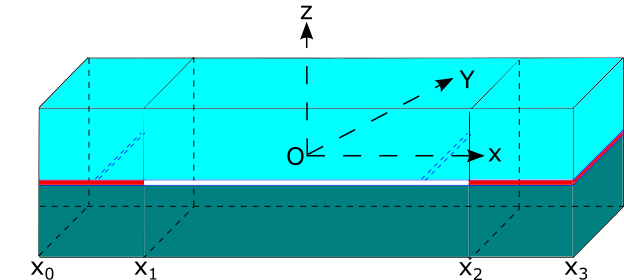} 
    \vspace{4ex}
  \caption{Bi-layer structure with an initial soft bonded region for $x_0<x<x_1$, a delaminated region for $x_1<x<x_2$ and a soft bonded region for  $x_2<x<x_3$.  We assume that the materials in both layers are similar,  that is, their material properties differ by $\mathcal{O}(\varepsilon)$.}
  \label{fig:SBFig} 
\end{figure}

The second case we consider is when we have a two layered waveguide with ``soft'' bonding between the layers. This is illustrated in Figure \ref{fig:SBFig}.
The longitudinal displacement in the bonded regions is described by the regularized non-dimensional equations
\begin{align}
     u^{(i)}_{tt}-u^{(i)}_{xx}&= \varepsilon \left[ -12u^{(i)}_{x}u^{(i)}_{xx} +2u^{(i)}_{ttxx} \right] \notag \\
     &~~~-\varepsilon\delta(u^{(i)}-w^{(i)}), \label{ucRB}
     \\ w^{(i)}_{tt}-c^2w^{(i)}_{xx}&= \varepsilon \left[ -12\alpha w^{(i)}_{x}w^{(i)}_{xx} + 2\beta w^{(i)}_{ttxx} \right] \notag \\
     &~~~+ \varepsilon\gamma(u^{(i)}-w^{(i)}),   \label{wcRB}
\end{align}
for $x_{i-1} < x < x_i$, while in the delaminated regions we have Boussinesq equations
\begin{align}
     u^{(i)}_{tt}-u^{(i)}_{xx}&= \varepsilon \left[-12u^{(i)}_{x}u^{(i)}_{xx} +2u^{(i)}_{ttxx} \right], \label{uBous}
     \\ w^{(i)}_{tt}-c^2w^{(i)}_{xx}&= \varepsilon \left[-12\alpha w^{(i)}_{x}w^{(i)}_{xx} +2\beta w^{(i)}_{ttxx} \right]. \label{wBous}
\end{align}
As with the perfectly bonded case, these equations are complemented with continuity conditions at the interfaces between the sections. We have continuity of longitudinal displacement 
\begin{align}
&u^{(i)} |_{x=x_i} = u^{(i+1)} |_{x=x_i}, \notag \\
&w^{(i)} |_{x=x_i} = w^{(i+1)} |_{x=x_i}, \label{cRBCont}
\end{align}
and continuity of normal stress
\begin{align}
&\sigma_u^{(i)}|_{x=x_i} = \sigma_u^{(i+1)}|_{x=x_i}, \notag \\
&\sigma_w^{(i)}|_{x=x_i} = \sigma_w^{(i+1)}|_{x=x_i},  \label{cRBCont2}
\end{align}
for $i=1,2$, where $\sigma_u$ and $\sigma_w$ are defined by \eqref{ucRB} and \eqref{wcRB} as
\begin{align*}
	u_{tt}^{(i)} &= \frac{\mathrm{d} \sigma_u^{(i)}}{\mathrm{d} x} - \delta(u^{(i)} - w^{(i)}), \\
	w_{tt}^{(i)} &= \frac{\mathrm{d} \sigma_w^{(i)}}{\mathrm{d} x} + \gamma(u^{(i)} - w^{(i)}),	
\end{align*}
respectively. We will consider the case here where the materials in the layers are similar, namely $c - 1 = \mathcal{O}(\varepsilon)$. We can construct a weakly-nonlinear solution to this system of equations, as was done in \cite{Khusnutdinova17}, however we cannot obtain any direct theoretical predictions from this approach as the derived coupled Ostrovsky equations are not solvable via the Inverse Scattering Transform. Therefore, we will explore this case numerically to determine a measure of delamination.

\section{Numerical results}
\label{sec:Numerical}
We now aim to use the derived weakly-nonlinear solution and the theoretical predictions of Section \ref{sec:Setup} to introduce a measure of the delamination length in terms of the change in wave structure.  In this section, we first demonstrate the effect of delamination on the transmitted soliton for both the perfectly bonded and soft bonded waveguides in Section \ref{sec:ScatteringNumerics}. We then introduce a measure of the delamination length for the perfectly bonded case in Section \ref{sec:PerfectBondNumerics}, and for the soft bonded case in Section \ref{sec:SoftBondNumerics}.  In both cases we consider how these measures scale with respect to the incident soliton in order to rapidly recompute results for a wide range of initial conditions. We will use the finite difference scheme from Ref.~\cite{Tranter19} to solve the original Boussinesq system and a semi-analytical method using a pseudospectral scheme for the derived KdV equations for the perfectly bonded case,  similar to the one used for coupled Ostrovsky equations in \cite{Khusnutdinova17}. In all cases our simulations will use a grid spacing of $\Delta x = 0.01$ and a time step of $\Delta t = 0.01$ for the finite-difference scheme. For the semi-analytical method we take $\Delta X = 5 \times 10^{-4}$ and $\Delta \xi = 0.1$, corresponding to $N = 131,072$. 

\subsection{Examples of scattering}\label{sec:ScatteringNumerics}
\begin{figure}[htbp] 
    \centering
  \begin{subfigure}[b]{0.6\linewidth}
   \centering
    \includegraphics[width=\linewidth]{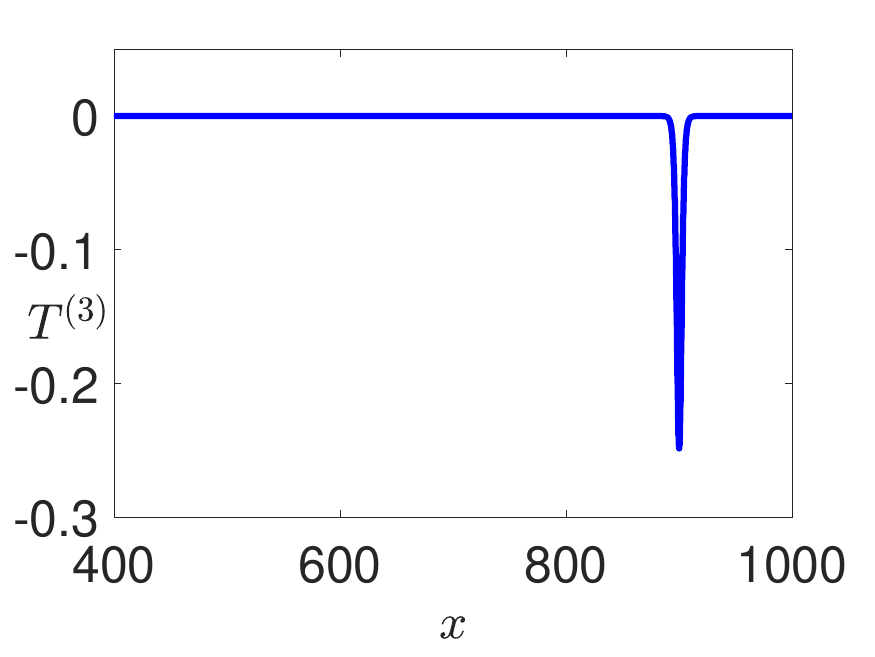} 
    \caption{$D = 0$ \label{fig:PBDelamA} } 
  \end{subfigure}
  \begin{subfigure}[b]{0.6\linewidth}
   \centering
    \includegraphics[width=\linewidth]{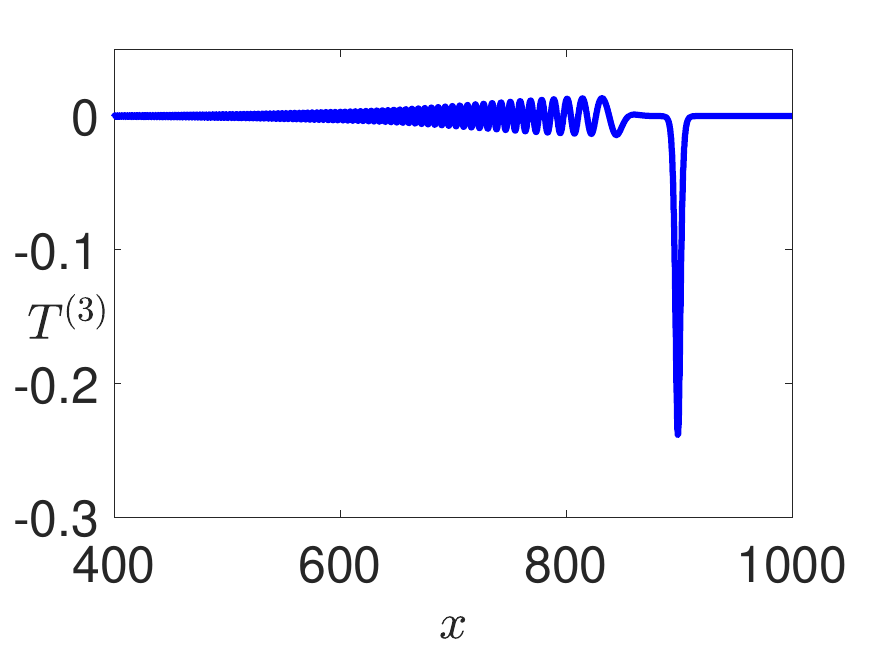} 
    \caption{$D = 50$ \label{fig:PBDelamB} } 
  \end{subfigure}
    \begin{subfigure}[b]{0.6\linewidth}
   \centering
    \includegraphics[width=\linewidth]{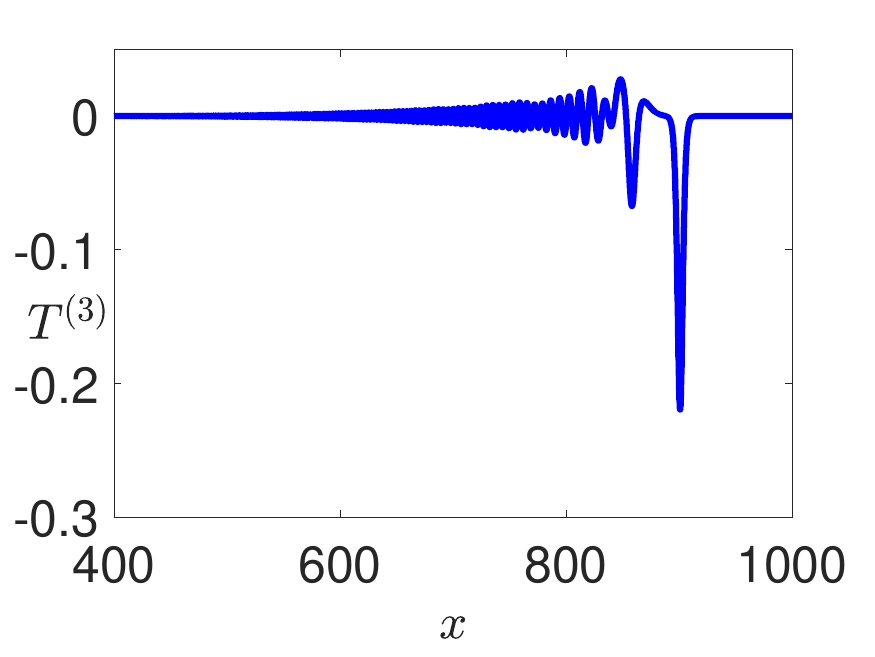} 
    \caption{$D = 300$ \label{fig:PBDelamC} } 
  \end{subfigure}
  \begin{subfigure}[b]{0.6\linewidth}
   \centering
    \includegraphics[width=\linewidth]{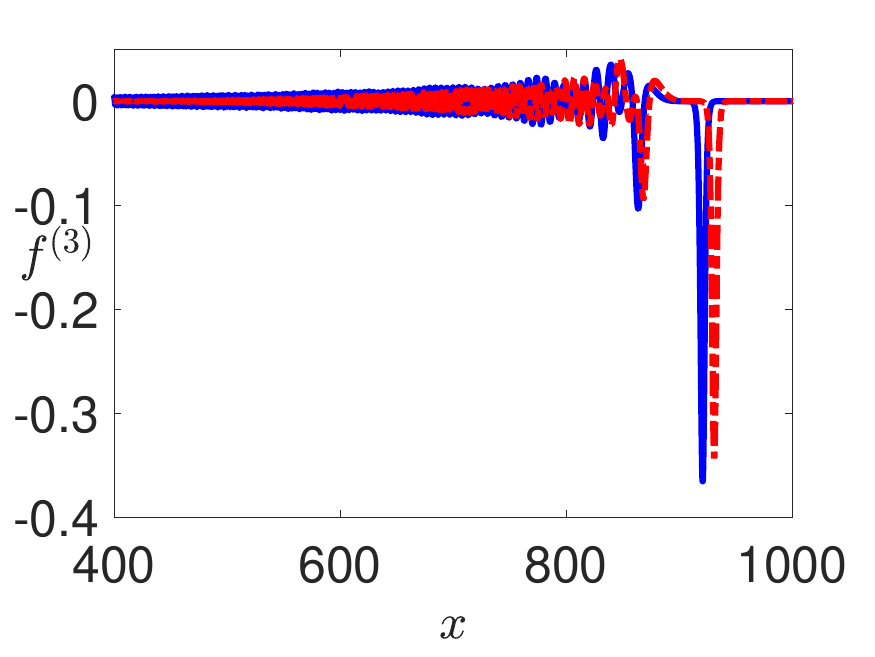} 
    \caption{Comparison of schemes for $D = 300$.} \label{fig:PBDelamD}
  \end{subfigure} 
  \caption{The solution at $t = 900$ in the final section of the perfectly bonded waveguide, for various delamination lengths, and comparison of the direct numerical (blue, solid line) and semi-analytical (red, dotted line) simulations. Parameters are $\varepsilon = 0.1$, $\textrm{FWHM} = 5.0$, $n = 2$ and $k = 2$. The finite-difference method uses a computational domain of $[-100, 1000]$ and for the semi-analytical pseudospectral method we have $N = 131,072$.}
  \label{fig:PBDelam} 
\end{figure}

Firstly, we demonstrate the effect of delamination on the propagation of an incident solitary wave, in both scenarios described in Section \ref{sec:Setup}. For the perfectly bonded case, let us assume a spatial domain $x \in [-100, 1000]$ and we have a delamination starting at $x = 0$ of length $D$. We present three cases: no delamination, a delamination of length $D = 50$ and a delamination of length $D = 300$. These results, as well as a comparison between the weakly-nonlinear solution and the directly computed solution, are shown in Figure \ref{fig:PBDelam}. We can see from Figure \ref{fig:PBDelamA} that, in the case of no delamination, the soliton propagates without change in shape or structure.  When delamination is introduced in figures \ref{fig:PBDelamB}  and \ref{fig:PBDelamC}, the soliton fissions into two solitons with accompanying dispersive radiation. Comparing the case of $D = 50$ to $D = 300$, we can see that the second soliton has become more distinct from the radiation and the first soliton has tended towards its theoretically predicted amplitude. Figure \ref{fig:PBDelamD} shows that there is a reasonable agreement between the direct numerical simulation and the semi-analytical method when $D = 300$,  with a slight phase shift and amplitude change. This will be reduced for smaller values of $\varepsilon$ (as we have constructed a series in increasing powers of $\varepsilon$), however we have qualitatively the same structure.

\begin{figure}[htbp] 
    \centering
  \begin{subfigure}[b]{0.6\linewidth}
   \centering
    \includegraphics[width=\linewidth]{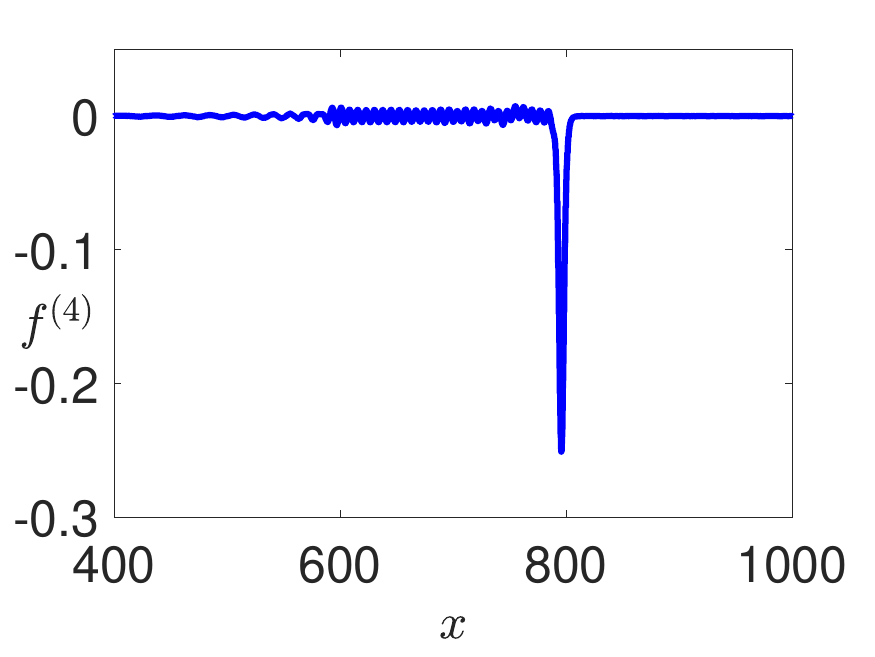} 
    \caption{$D = 0$ \label{fig:SBDelamA} } 
  \end{subfigure}
  \hspace{0.05\linewidth}
  \begin{subfigure}[b]{0.6\linewidth}
   \centering
    \includegraphics[width=\linewidth]{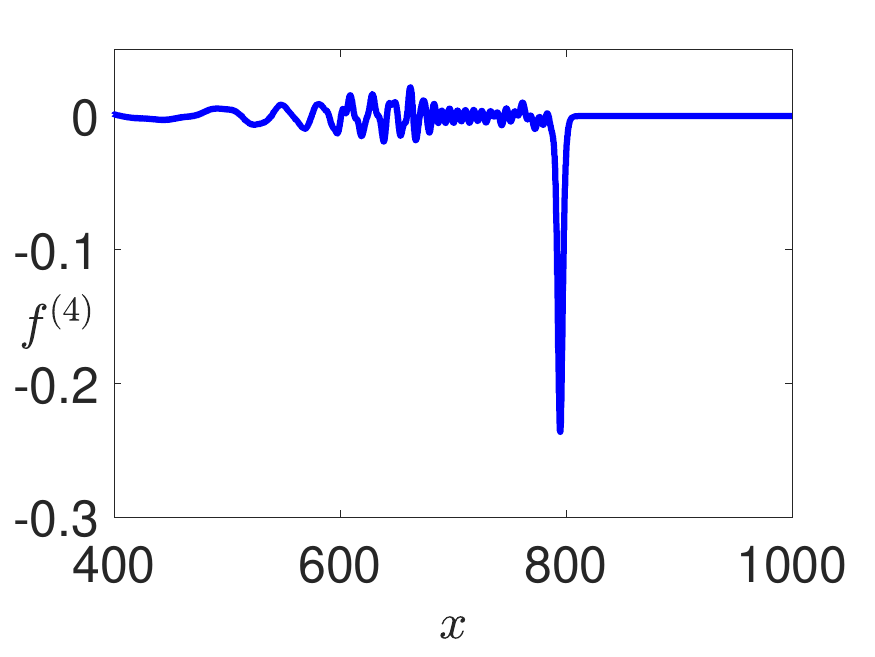} 
    \caption{$D = 100$\label{fig:SBDelamB}} 
  \end{subfigure}
    \begin{subfigure}[b]{0.6\linewidth}
   \centering
    \includegraphics[width=\linewidth]{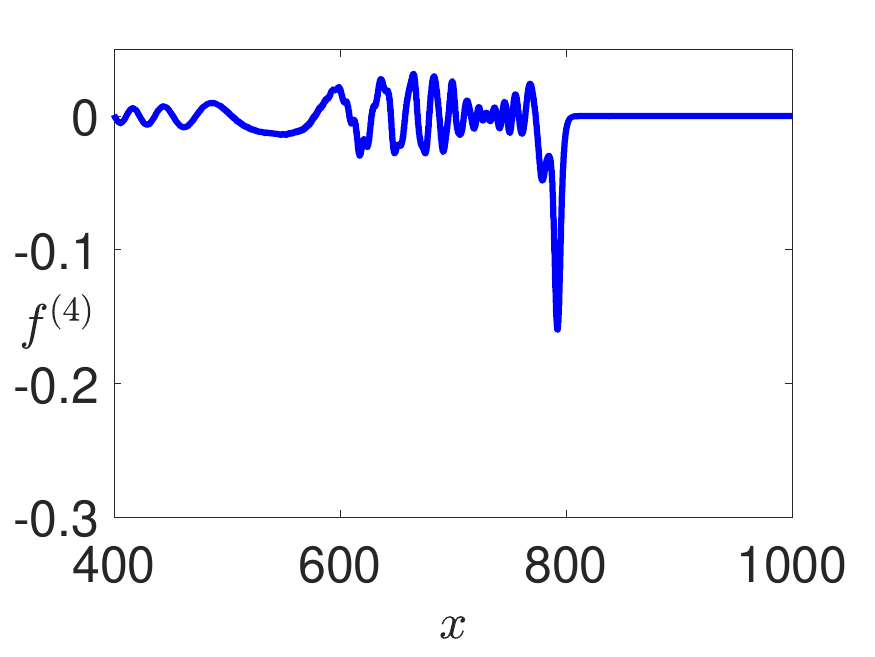} 
    \caption{$D = 300$\label{fig:SBDelamC}} 
  \end{subfigure}
  \caption{The solution at $t = 900$ in the final section of the upper layer of the soft bonded waveguide, for various delamination lengths. Parameters are $\varepsilon = 0.05$, $\textrm{FWHM} = 5.0$, $c = 1.025$, $\alpha = \beta = 1.05$, $\delta = \gamma = 1$. The finite-difference method uses a computational domain of $[-500, 1000]$.}
  \label{fig:SBDelam} 
\end{figure}%

Similarly, we consider the case of soft bonding and an incident soliton. We assume a spatial domain $x \in [-500, 1000]$, with a homogeneous (delaminated) section for $x \in [-500, -400]$ to generate the same wave in both layers, bonded sections $x \in [-400, 0]$ and $x \in [D, 1000]$, with a delaminated region for $x \in [0, D]$. We again present three cases including when there is no delamination,  a shorter delamination length of $D = 100$ and a larger delamination with $D = 300$. The results are shown in Figure \ref{fig:SBDelam} for the upper layer, where the lower layer is qualitatively similar. In the case of no delamination shown in Figure \ref{fig:SBDelamA} we have a solitary wave with a one-sided oscillatory tail, known as a \emph{radiating solitary wave}. As the delamination length increases, the solitary wave begins to lose amplitude and expel energy into its tail through an exchange of energy between the layers. These are clear signs of the presence of delamination in figures \ref{fig:SBDelamB} and \ref{fig:SBDelamC}, with structural changes that can be detected and the decay in amplitude can be quantified to give a measure of the length of delamination. Note that in this case there is no comparison between the simulation schemes since the semi-analytical scheme is not applicable for the soft bonded case.

\subsection{Measure of delamination length for perfect bonding}\label{sec:PerfectBondNumerics}
We now expand upon the observations from the previous subsection by introducing a measure of delamination based upon the theoretical predictions from Section \ref{sec:Setup}.  We then generalise this measure to apply for different incident soliton widths. If we assume that the solitons are well-separated, representing the case of infinite delamination, the amplitude of the soliton can be found from the IST. We have
\begin{align}
    A_3 &= A_1 k_2^2 k_3^2, \quad k_2 = \frac{1}{2} \left( \sqrt{1 + \frac{8}{\beta_2}} - 1 \right), \notag \\
    k_3 &= \frac{1}{2} \left( \sqrt{1 + 8\beta_2} - 1 \right),
\end{align}
where $A_1$ is the amplitude of the incident soliton, $A_3$ is the amplitude of the lead soliton in the second bonded region, and $k_2$, $k_3$ are the eigenvalues corresponding to the lead soliton amplitude in the second and third regions, as determined by the IST. As the delamination length is reduced, the amplitude in the third region will tend towards the initial amplitude, $A_1$. 

Denoting the calculated numerical solution as $A_{\mathrm{num}}$ from the simulation, we introduce a measure of the amplitude of the lead soliton in the third section of the bar in comparison to the incident soliton as
\begin{equation}
    \sigma=\frac{A_{\mathrm{num}}-A_1}{A_3-A_1} \times 100.
\end{equation}
This corresponds with the measure used in \cite{Tranter19}, where it was assumed that $\mathrm{FWHM} = 5$. We now compute the solution using the semi-analytical pseudospectral scheme for a wide range of values of FWHM with the aim of determining a general rule for the delamination length. The delamination length is chosen to be $D \in [0, 300]$ and we measure the delamination in multiples of FWHM.

\begin{figure}[htbp] 
    \centering
  \begin{subfigure}[b]{0.7\linewidth}
    \centering
    \includegraphics[width=\linewidth]{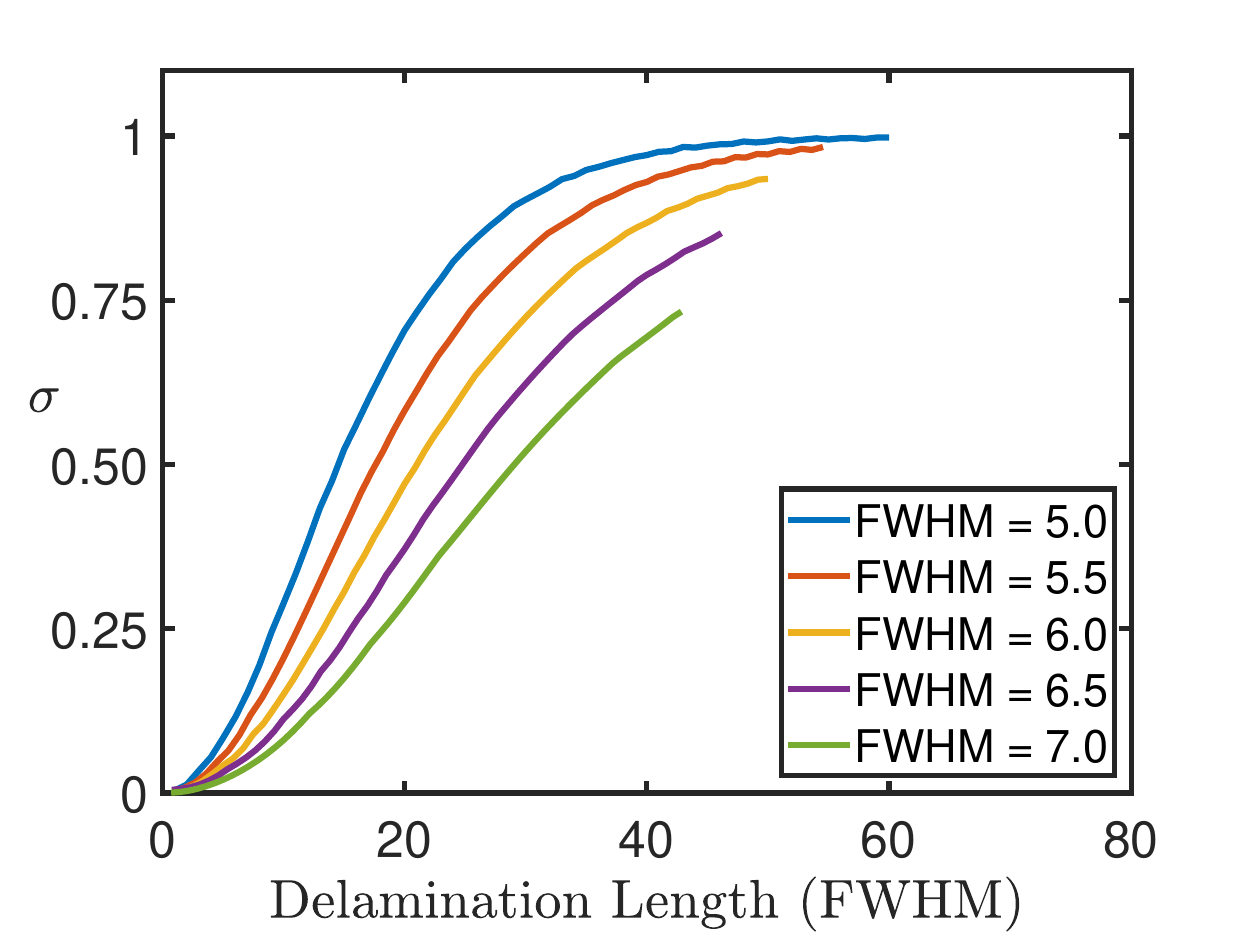} 
    \caption{Delamination measure for $n=3$, $k=3$.} 
    \label{fig:PBUnscaledn3k3} 
  \end{subfigure}
  \begin{subfigure}[b]{0.7\linewidth}
    \centering
    \includegraphics[width=\linewidth]{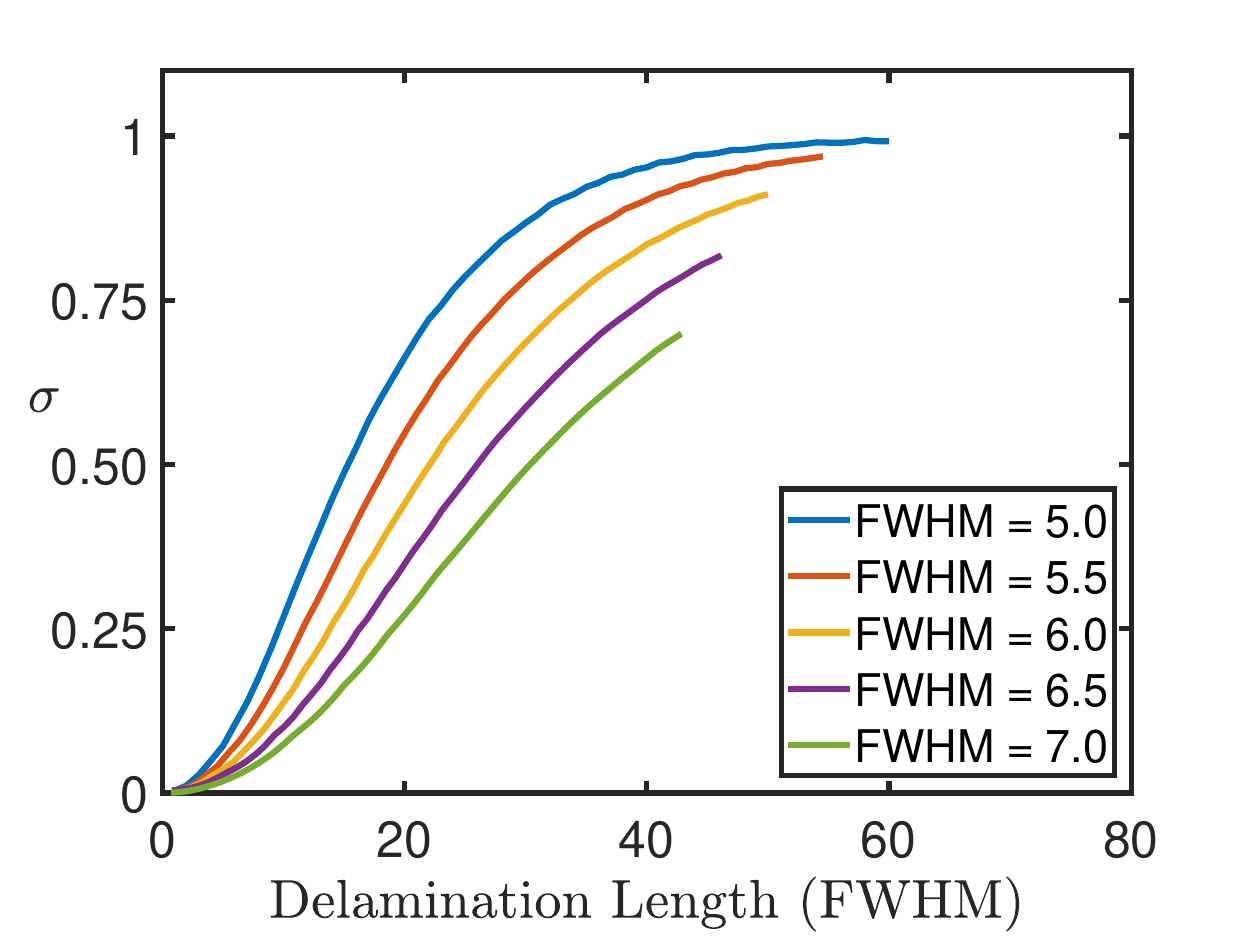} 
    \caption{Delamination measure for $n=4$, $k=3$.} 
    \label{fig:PBUnscaledn4k3} 
  \end{subfigure}  
  \caption{Plots of the change in amplitude of the lead transmitted soliton, in comparison to the incident soliton and theoretical predictions, as measured by $\sigma$, for various values of FWHM. Here we take $\varepsilon = 0.1$.}
  \label{fig:PBUnscaled} 
\end{figure}

The results are plotted in Figure \ref{fig:PBUnscaledn3k3} for $n = 3$, $k = 3$, and in Figure \ref{fig:PBUnscaledn4k3} for $n = 4$ and $k = 3$. We can see that, as the delamination length increases, the measure $\sigma$ increases until it tends to a value of $1$, and this behaviour is replicated for different values of FWHM. For larger FWHM it may not reach this limit for the chosen delamination length. We can also see a similar behaviour for different values of $n$ and $k$.

\begin{figure}[htbp] 
   \centering
   \begin{subfigure}[b]{0.7\linewidth}
    \includegraphics[width=\linewidth]{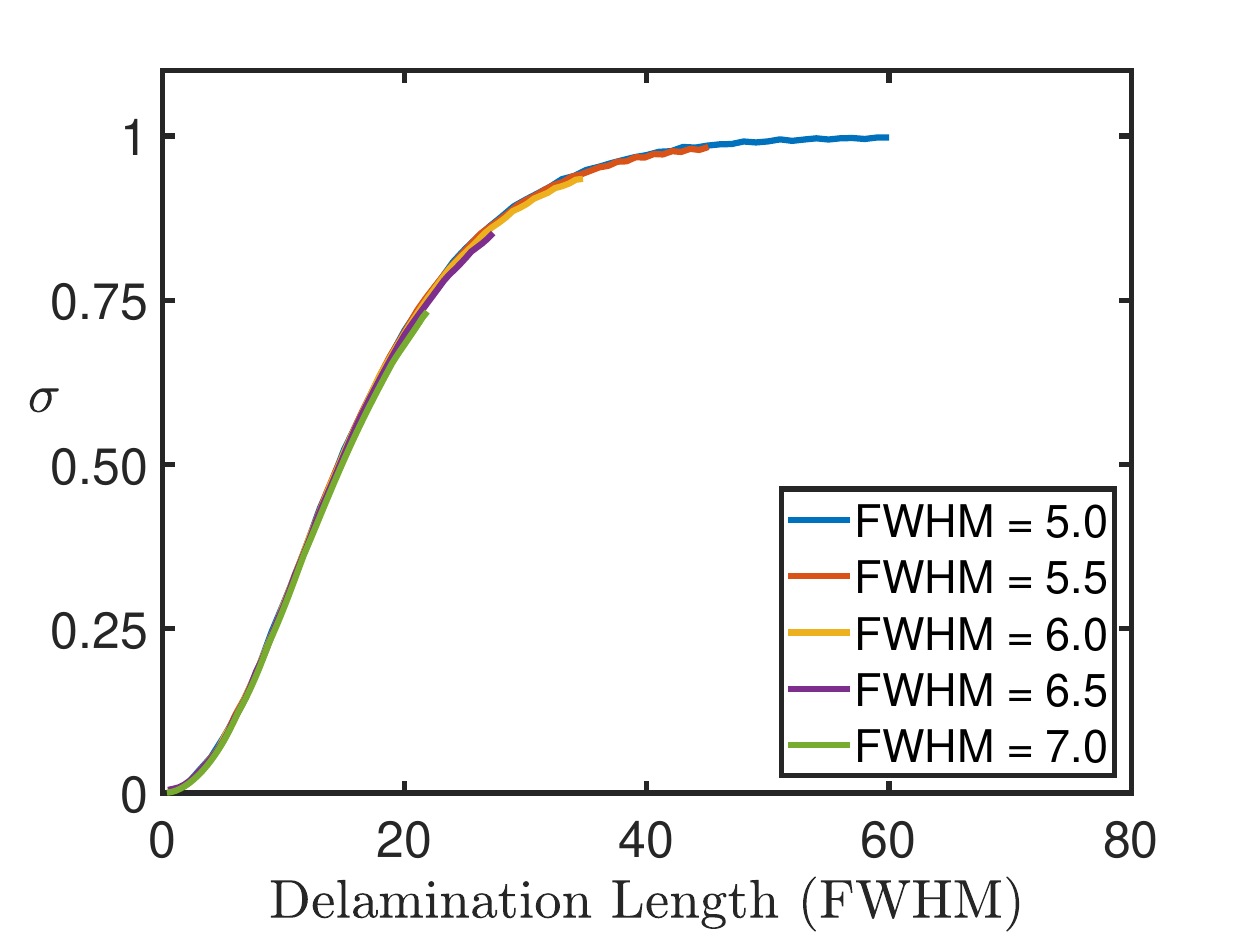} 
    \caption{Scaled delamination curve $n=3$, $k=3$.} 
    \label{fig:PBScaledn3k3} 
    \vspace{4ex}
  \end{subfigure}
  \begin{subfigure}[b]{0.7\linewidth}
   \centering
    \includegraphics[width=\linewidth]{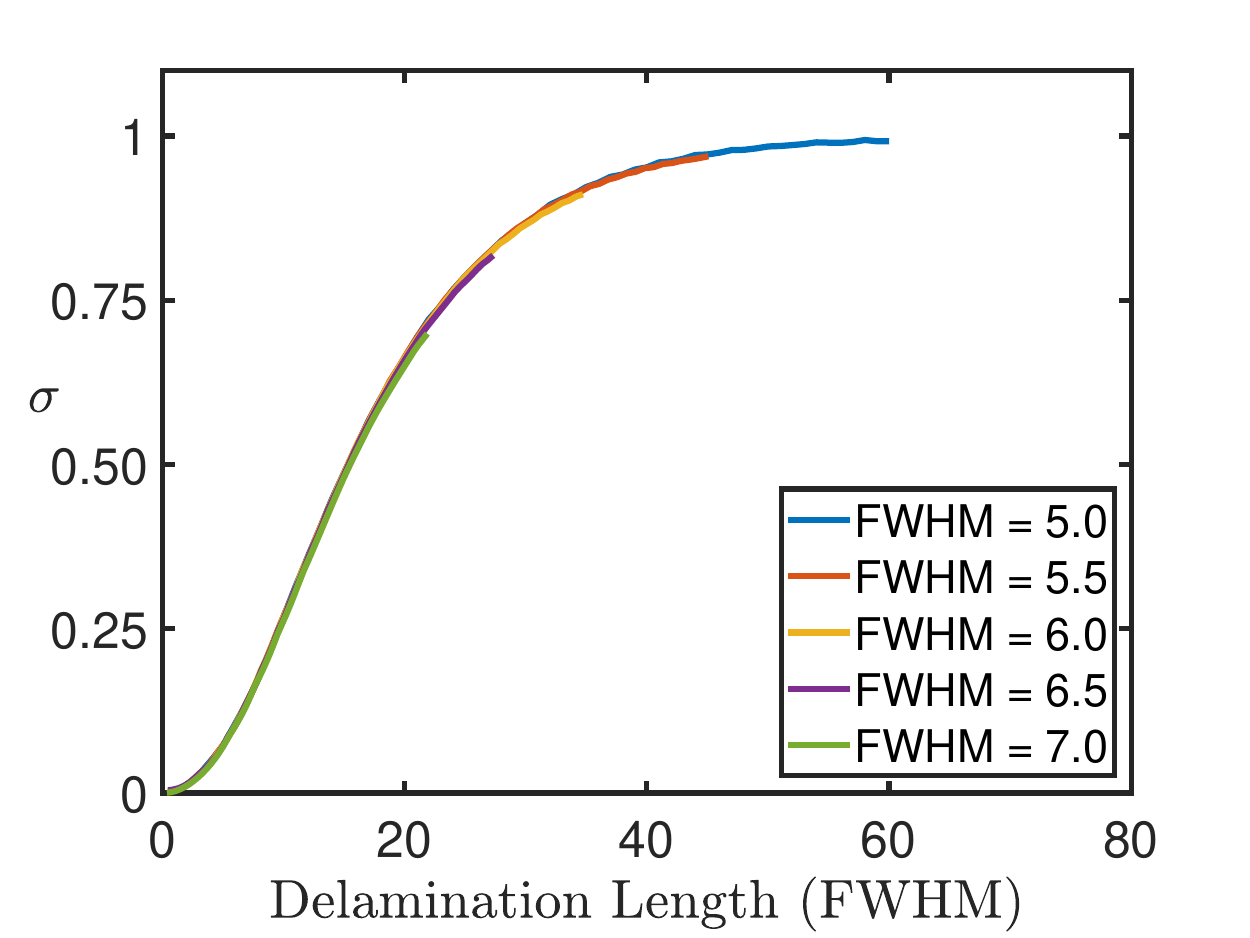} 
    \caption{Scaled delamination curve $n=4$, $k=3$.} 
    \label{fig:PBScaledn4k3} 
    \vspace{4ex}
  \end{subfigure} 
  \caption{Plots of the scaled delamination curves for the change in amplitude of the lead transmitted soliton, in comparison to the incident soliton and theoretical predictions, as measured by $\sigma$. Here we take $\varepsilon = 0.1$. }
  \label{fig:PBScaled} 
\end{figure}

To generalise this approach, we consider the equation for the phase speed $v$ in terms of FWHM \eqref{FWHM}.  We can see that $v$ in inversely proportional to the square of FWHM. Thus, fixing our reference as $\mathrm{FWHM} = 5$, we let $\sigma$ be a function of delamination length, parametrised by FWHM, and we introduce the scaling
\begin{equation}
    \tilde{\sigma}(D;\mathrm{FWHM}) = \frac{\mathrm{FWHM}^2}{25} \sigma(D;\mathrm{FWHM}).
    \label{FWHMScale}
\end{equation}
The resulting plots are shown in Figure \ref{fig:PBScaledn3k3} for $n=3$, $k=3$, and in Figure \ref{fig:PBScaledn4k3} for $n=4$, $k=3$. We see that the scaled versions are very closely aligned, with the only disagreement stemming from the restriction on delamination length for larger values of FWHM. Therefore, after computing one curve for the smallest value of FWHM, we can reproduce all subsequent curves for any value of FWHM. This allows for the highly efficient computation of the delamination curves and for a wide range of experiments with incident waves of different amplitude.

\subsection{Finite delamination with soft bonding}\label{sec:SoftBondNumerics}
We now consider the soft bonded case outlined in Section \ref{sec:ImperfBond}.  This case was also studied in Ref.~\cite{Khusnutdinova17},  but for only one value of FWHM.  In this work we compute the solution for a wide range of FWHM, using the finite-difference scheme in \cite{Tranter19}. The constructed weakly-nonlinear solution consists of coupled Ostrovsky equations, in contrast to the KdV equations in the previous case \cite{Khusnutdinova17}. Therefore we cannot use the IST to predict the amplitude of the waves in the bonded regions as coupled Ostrovsky equations are not integrable via the IST. The incident soliton in this case evolves into a radiating solitary wave, that is a solitary wave with a one-sided oscillatory tail.

\begin{figure}[htbp] 
    \centering
  \begin{subfigure}[b]{0.7\linewidth}
   \centering
    \includegraphics[width=\linewidth]{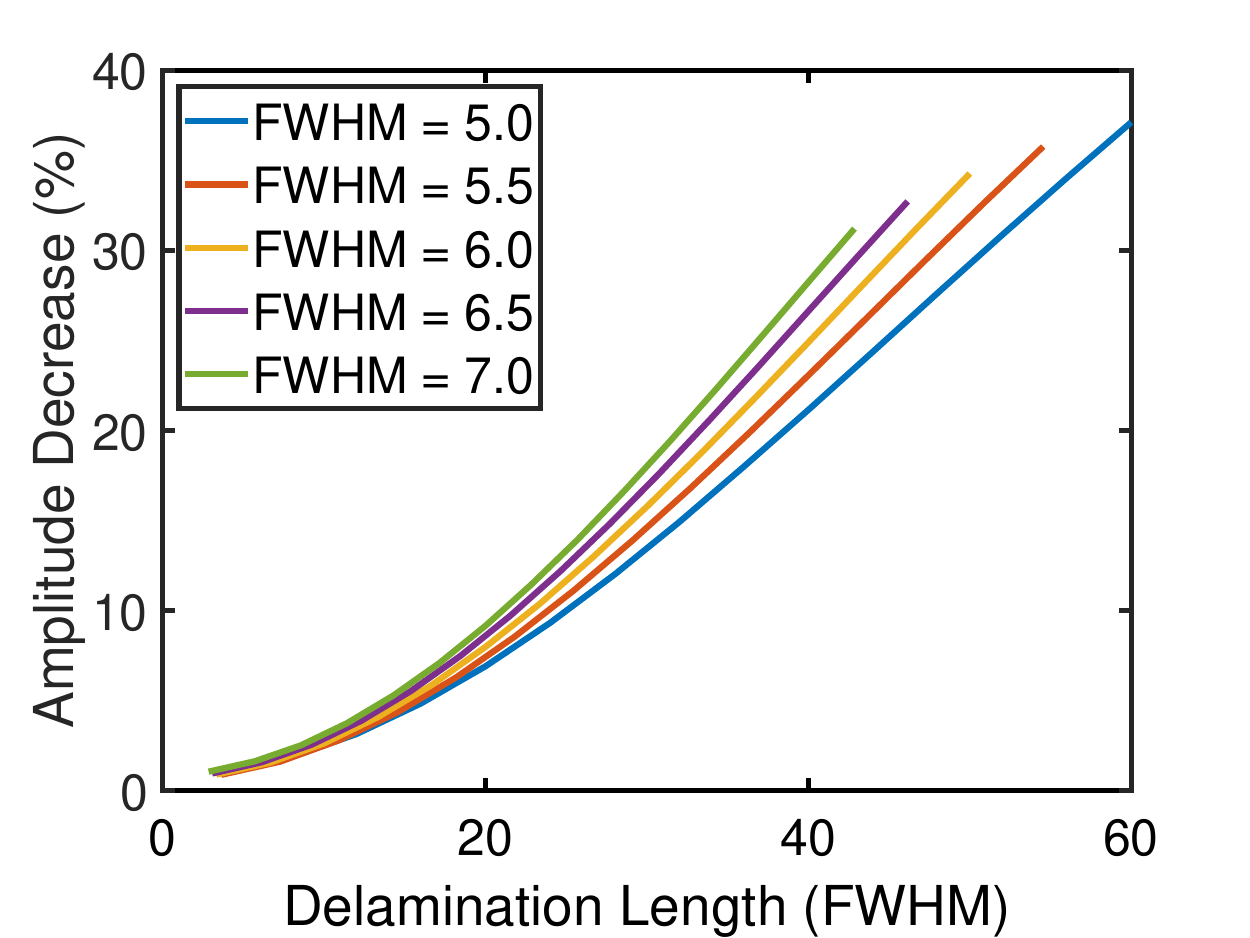} 
    \caption{Amplitude change in upper layer} 
    \vspace{4ex}
  \end{subfigure}
  \begin{subfigure}[b]{0.7\linewidth}
   \centering
    \includegraphics[width=\linewidth]{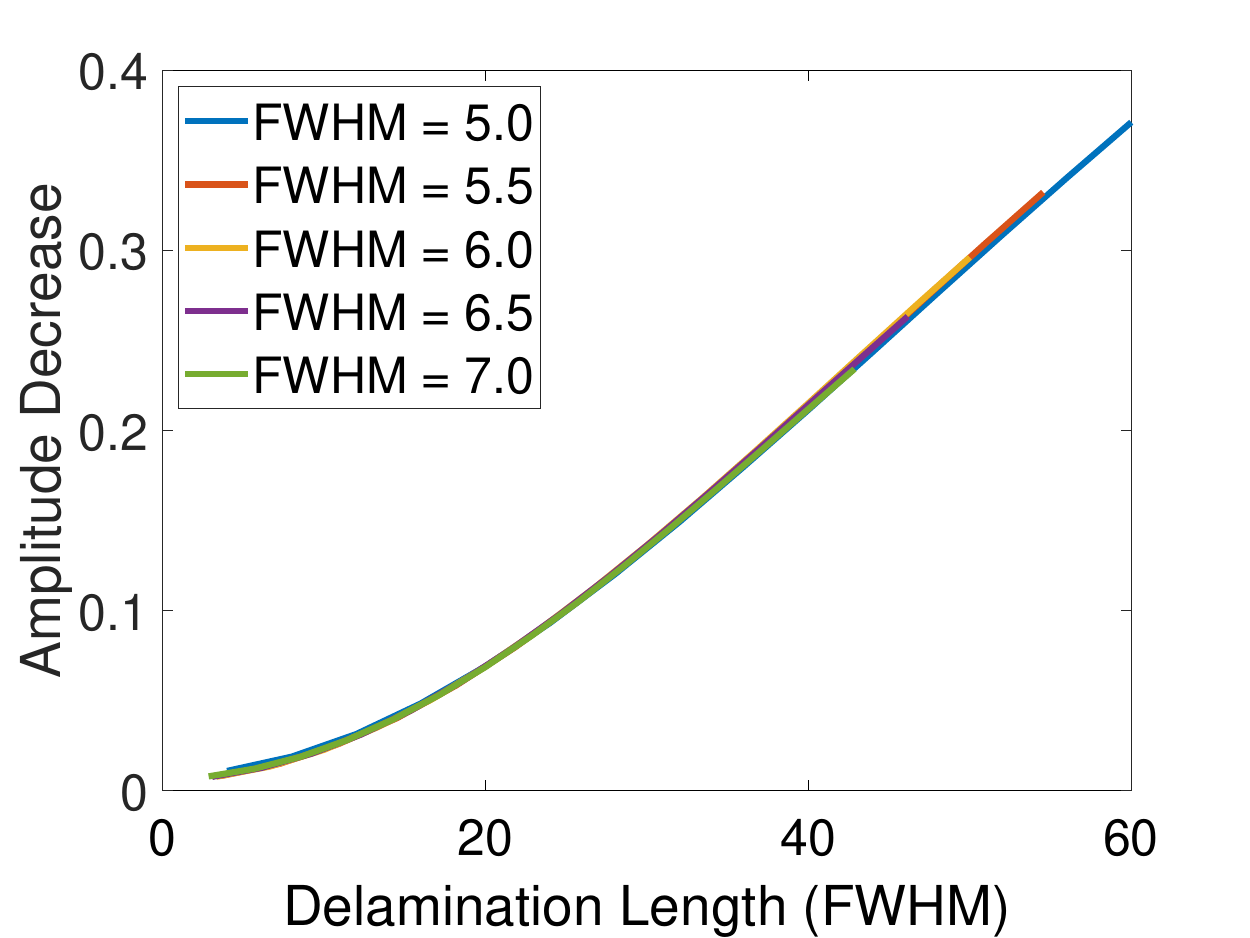} 
    \caption{Amplitude change in lower layer} 
    \vspace{4ex}
  \end{subfigure} 
  \caption{Change in amplitude of the radiating solitary wave in the soft bonded structure, for various delamination lengths and values of FWHM. Parameters are $\varepsilon = 0.05$, $\delta = \gamma = 1$. }
  \label{fig:SBFWHM} 
\end{figure}

To determine the change in amplitude, as a measure of the delamination length, we denote the amplitude of the soliton or wave packet in each region as $A_L$, where $L$ is the region index, and we introduce the measure
\begin{equation}
    \zeta = \frac{|A_1-A_3|}{A_1} \times 100.
\end{equation}
Figure \ref{fig:SBFWHM}(a) presents the results for $\zeta$, computed over a wide range of FWHM values, for the upper layer. A similar agreement is seen for the lower layer, but the results are omitted for brevity.  Following the idea from the perfectly bonded case, we introduce a quadratic scaling. We choose a reference FWHM value (in this case we choose our lowest value of FWHM, namely $\mathrm{FWHM} = 5$) and then calculate a scaling of the form
\begin{equation}
	\tilde{\zeta} = \frac{\zeta}{a + b\ \overline{\mathrm{FWHM}} + c\ \overline{\mathrm{FWHM}}^2},
	\label{SBScaling}
\end{equation}
where we have introduced
\begin{equation*}
	\overline{\mathrm{FWHM}} = \frac{\mathrm{FWHM}}{5},
\end{equation*}
and $a$, $b$, $c$ are constants to be found that satisfy the relationship $a + b + c = 1$. The results for $a = 0.49$, $b = 0.28$, $c = 0.23$ are plotted in Figure \ref{fig:SBFWHM}(b) and we can see a good agreement across a range of values of FWHM. However, this fitting is done by careful choice of parameters rather than theoretical prediction, as for the previous case. The agreement begins to worsen slightly after a delamination of 40 units of FWHM, which corresponds to a minimum of 200 in nondimensional units, however overall the agreement is still good.

We now summarise the scaling used to convert our nondimenesional variables to dimensional material parameters that can be compared to experimental data, in order to confirm whether a delamination length of 200 is reasonable.  Referring to the dimensional form of the DDE in \eqref{DDE}, with parameters \eqref{DDEParam}, we introduce the scaling to nondimensional form via
\begin{equation}
	\tilde{x} = \frac{x}{X}, \quad \tilde{f} = \frac{f}{F}, \quad \tilde{t} = \frac{t}{T},
	\label{MatScaling}
\end{equation}
where
\begin{align}
	X &= \sqrt{\frac{J \nu^2}{2\varepsilon \sigma c^2} \left( c^2 - c_1^2 \right)}, \quad T = \frac{X}{c}, \notag \\
	F &= -\frac{12 \varepsilon c^2 \rho}{\beta} X.
\end{align}
We can therefore find the corresponding material length given the nondimensional length. Let us assume a PMMA bar of 10mm $\times$ 10mm cross-section, then using the parameters for PMMA from \cite{Dreiden12} we find that, for $\varepsilon = 0.1$, a delamination length of $x = 200$ in nondimensional units corresponds to a length of approximately $x = 520$mm, which is significant given the experimental materials are usually around $600$mm long in total. Therefore, restricting our considerations to values of delamination less than 200 nondimensional units is reasonable in the context of practical applications.

\section{Conclusion}
\label{sec:Conc}
In this paper we have considered the scattering of a bulk strain solitary wave in a delaminated bi-layer with either perfect or soft bonding between the layers. The longitudinal displacements are modelled by either Boussinesq equations (perfectly bonded or delaminated sections) or coupled Boussinesq equations (soft bonded sections), with continuity conditions on the interface. Incident solitary waves undergo fission in delaminated regions in the perfectly bonded structure, providing a clear indicator of delamination.

We construct theoretical estimates for the amplitude of the solitons after a delaminated region, using the Inverse Scattering Transform. A measure is introduced using the theoretical and observed values to predict the delamination length based upon amplitude changes. This is then extended for incident waves of different Full Width at Half Magnitude, and a quadratic scaling is introduced and verified by numerical results. The significance of this result is that we now only need to compute a single curve in order to perform a wide range of experiments, which significantly reduces computation times and allows for further experiments (with different incident solitons) to be performed rapidly. This was confirmed for various configurations of the waveguide.

In the case of a soft bonded waveguide with delamination, theoretical estimates cannot be derived using the Inverse Scattering Transform.  In this case we resort to direct computation of the solution and a comparison between the amplitude after delamination and the corresponding amplitude for the non-delaminated case.  A similar quadratic scaling can be found,  which has a good agreement up to a delamination length of 200 in nondimensional units or 520mm in physical units. This is consistent for both layers of the waveguide. 

This work facilitates experimentation with a wide range of initial condition parameters, and provides a framework for detecting delamination in perfectly bonded and soft bonded waveguides with similar materials in the layers. The case with distinctly different materials in the layers is more complex, and some preliminary studies have been conducted into quantifying delamination \cite{Tamber22}.

\backmatter

\section*{Declarations}

\begin{itemize}
\item Funding:
Jagdeep S.~Tamber would like to thank Nottingham Trent University for funding through their PhD studentship scheme. 
\item Competing interests:
The authors have no relevant financial or non-financial interests to disclose.
\item Ethics approval:
Not applicable.
\item Consent to participate:
Not applicable.
\item Consent for publication:
Not applicable.
\item Availability of data, code and materials:
The datasets generated during this study can be reproduced using equations throughout the paper and the cited numerical methods. The codes used to generate the datasets are available from the corresponding author on reasonable request.
\end{itemize}

\bibliography{Research}

\end{document}